# Towards Learning Geometric Transformations through Play:

An AR-powered approach

ZOHREH, ZSH, SHAGHAGHIAN

Department of Architecture, Texas A&M University, College Station, Texas, USA, zohreh-sh@tamu.edu

WEI, WY, YAN

Department of Architecture, Texas A&M University, wyan@tamu.edu

DEZHEN, DS, SONG

Department of Computer Science and Engineering, Texas A&M University, dzsong@cs.tamu.edu

Despite the excessive developments of architectural parametric platforms, parametric design is often interpreted as an architectural style rather than a computational method. Also, the problem is still a lack of knowledge and skill about the technical application of parametric design in architectural modelling. Students often dive into utilizing complex digital modelling without having a competent pedagogical context to learn algorithmic thinking and the corresponding logic behind digital and parametric modelling. The insufficient skills and superficial knowledge often result in utilizing the modelling software through trial and error, not taking full advantage of what it has to offer. Geometric transformations as the fundamental functions of parametric modelling is explored in this study to anchor learning essential components in parametric modelling. Students need to understand the differences between variables, parameters, functions and their relations. Fologram, an Augmented Reality tool, is utilized in this study to learn geometric transformation and its components in an intuitive way. A LEGO set is used as an editable physical model to improve spatial skill through hand movement beside an instant feedback in the physical environment.

## 1 INTRODUCTION

Due to the benefits of digital modelling in rapid prototyping, simulation, and alternative design generation, most teachers aim to rapidly teach 3D modelling software to students in the early stages of design. However, a competent context for learning parametric modelling concepts and essentials of "digital design thinking" is rarely provided in architectural educational systems. Indeed, a general understanding of fundamental programming concepts behind computational design modelling may help students better understand the logic behind modelling software. Students are expected to identify and utilize modelling techniques without a proper understanding of digital design components named as variables (as the properties of the geometry), parameters (as the members of the function family), and functions (as the mapping operations). Superficial knowledge of digital design thinking often results in learning computational modelling through trial-and-error.

Learning geometric transformations as one of the fundamental components of parametric modelling is essential for effective development of 3D modelling in computational design [1]. Analyzing complex architectonic forms usually requires a high level of visualization and mental modelling skills and transformation knowledge to decompose multiple transformations mentally. Such skills become specifically critical in applying parametric modelling techniques, which usually starts from simple geometry but often results in a complicated geometric form through multiple transformations such as Rotation, Translation, Dilation, etc. However, the difficulty of learning geometric transformation and spatial-based problems is acknowledged clearly in the literature [2][3]. Understanding geometric transformation as mappings and functions rather than simple motions improves students' perception of transformation concepts [4]. Hence, It can help them in mental prediction and analysis of geometric transformation beyond verified action [2]. Such advanced level of reasoning helps students understand the logic behind parametric / digital modelling techniques and may assist them in utilizing parametric modelling techniques more efficiently and professionally. Meanwhile, the significance of traditional methods including physical modelling and hand drawing is well acknowledged in perceiving geometry and geometric relations for architectural and engineering students [5][6]. However, the quick switch towards digital modelling, because of their enormous privilege, may result in less exploitation of the potential benefits of physical models in cognitive learning, mental visualization, design innovation, and problem-solving.

Augmented Reality (AR) technology as a mediator tool, capable of integrating digital environment with physical real-world, can provide a spatial experiment to support embodied learning and virtual augmentation of information and abstract, "putting answers right where the questions are" [7]. The inherent capability of AR in superimposing the virtual information, including digital modelling and computer graphics (e.g., arrows, tags, highlighting, etc.) in a 3D space may help students perceive geometric transformations as mappings and functions beyond motions while exploring corresponding parameters in an integrated spatial scenario. In the meantime, the physical experiment in a 3D environment could play as a feedback besides improving spatial visualization skill through playing with a tangible manipulative.This study explores the educational potential of AR with Fologram in learning geometric transformations and  digital design thinking before diving into complex parametric modelling methods. Fologram is an AR tool capable of synching parametric modelling with AR environment [8]. This study intends to break down the process of basic geometric transformations into their function components in a learning AR environment with real-time feedback using physical model. In this environment, students can play with multiple parameters of the transformation functions to transform geometry's variables and follow the augmented result in the AR environment through tracing the transformation with the physical model in the physical environment. A LEGO model, as an editable tangible interface, is proposed in this study to keep track of the digital parametric changes to fortify learning impact and spatial skills while supporting feedback. This methodology is expected to improve student's understanding of geometric transformations and their corresponding components (variables, parameters, and operation functions) in a "Learning Through Play" environment.

## 2  RELATED WORK

The existing computer-assisted methods in enhancing students' learning of geometry and spatial transformations confirm the benefits of digital models' dynamic features. These studies acknowledge that exploring geometry and geometric relations from different perspectives helps students better understand geometry and improve their spatial skills, especially mental rotation, and spatial visualization [9]–[11]. However, many students still face challenges in solving geometric problems and mostly rely on a trial-and-error method [12].

Parametric design techniques may allow the students to understand geometry and geometric transformations by learning the impact of variables and parameters on their geometric design process. This level of reasoning can help them in improving design creativity and reframing design problems [13]. However, in many cases, at the end of the design studio students cannot identify their process in arriving to their final geometric solution since their work is mostly done through trial-and-error rather than a higher level of reasoning [14]. Longitudinal-experimental studies reveal that working with 3D modelling alone may not improve students' spatial skills and geometric perception; thus, traditional methods (e.g., hand drawing and physical modelling) are still required [15][16]. Physical modelling on the other hand, as a "versatile" tool for the designers and architects to express their thoughts in the design process is the most realistic and tangible 3D means to explore design thoughts and presentations [6]. Studies have acknowledged the benefits of such tangible tools in improving spatial visualization skills [17]. Psychological studies argue that physical interaction encourages students' epistemic actions [18]. Such skills help them in forming embodied metaphors of abstracts and internalizing the information that could enhance memory retrieval [18].

AR has been introduced in literature as the extension of Virtual Reality (VR) that let users take advantage of the synthetic environment while having immediate interaction with their physical surroundings [19]. AR as a tool supporting situated cognition environment can provide a spatial experiment where learning and applying knowledge occur at the same time in the same place [18]. AR has been widely used in STEM (Science, Technology, Engineering, and Mathematics) education such as physics, mathematics, and chemistry to support real-world perception of abstract information  to ease the learning process. The results show that AR has positive impact on spatial visualization skill [20], understating of mathematical concepts [21], learning performance, and higher knowledge gain [22][23]. In a recent experimental study conducted by Fidan et al. 2019, the impact of AR system integrated with Problem Based Learning (PBL) technique is investigated to assess students' learning achievement and knowledge retention in learning physics and natural phenomena [22]. The results reveal that AR technique not only fortifies PBL but also facilitates transfer of knowledge to real-life problems as well as promoting students' engagement in learning the targeted subject [22]. The application of AR in architecture and architectural education mostly has focused on visualizing building components and information [24][25], building energy performance and lighting simulation [26][27], and interior design augmentation [28][29]. Very few projects have recently started to explore integrating parametric components with parametric modelling through AR intervention [8], or a complete VR environment via VR tools [30]; however, none of the studies has explored the educational potential of AR in learning parametric modeling.

## 2.1  Reflective feedback in education

AR provides a context where hands-on experiences could take place in a physical environment along with digital information. This feature could potentially support a real-time feedback through physical realization of digital interaction in the real-world environment.

Extensive research on the importance of self-regulated learning (SRL) and the role of self-regulation mechanisms in learning enhancement and compensation of individual differences suggests that proactive learning  has a significant impact on students' academic success [31]. These studies demonstrate that SRL could be accomplished through constructive feedback as a learning tool [32].  Meanwhile the feedback are only useful if students are afforded with opportunity to integrate it with their prior knowledge [33]. Studies reflect that heterogeneous student populations, lack of resources, and population increase in higher education result in deficient constructive feedback to provide students with the opportunity to enhance their learning [32][34]. Most students require detailed feedback on learning the targeted subject; however, they mostly value their grades rather than feedbacks and comments to improve their learning process [35]. This problem may become more critical in architectural education in learning digital modeling techniques that is utilized only as a tool to realize the design goal and less focus is conducted in learning the modeling tool professionally. Lack of constructive feedback in modeling software could be a reason why students are learning through trial-and-error. Constructive feedbacks help students with self-regulation process, stimulate their motivation, and improve students self-esteem [31].

## 3  METHODOLOGY

The study proposes three educational aids: 1) Grasshopper as a parametric modeling tool, 2) Fologram, an AR tool capable of synching parametric modeling with AR environment to augment abstract information, and 3) a tangible LEGO model (#5891) as an editable physical manipulative. The digital model is an FBX model imported into Grasshopper. Other digital models could also be imported to the program and used as a model target in Fologram; however, customizations may be required to realize the prototype's tasks. QR code is utilized in AR registration to superimpose digital model and abstractions on the physical environment.

The study uses the capability of Fologram to synchronize the corresponding parameters of different geometric transformations such as translation and rotation in Grasshopper to help students understand geometric transformations as mappings and functions rather than simple motions. The corresponding parameters, synchronized to the AR app, could be modified by the player using sliders and toggle buttons to apply desired transformations. The transformations could be applied to parts of the LEGO model that are transformable in the physical environment. The LEGO model #5891 is a house model with editable parts such as a movable attic and rotatable doors/windows. Graphical abstractions such as coordinate system, rotation angles, translation vectors, and distance notations are visualized in the AR environment to help students understand different parameters of transformation functions in an intuitive way and track them through actual motion in the physical environment. Two prototypes are conducted for two common types of transformations: translation and rotation.

### 3.1  Prototype1

In this prototype, Grasshopper component for translation is utilized with two inputs including targeted geometry as the variable and translating vector as the function parameter. The translation vector is broken into three sliders representing translation values in each coordinate axis (x, y, and z). Although variables are the points, lines and surfaces of a corresponding geometry, to simplify the process the attic part of the LEGO model which is movable in the real LEGO is utilized as the variable to apply translation.

Students can play with the function parameters and observe the impact of parameter changes on the corresponding variable. The translation vector besides a notation which demonstrates the translation amount will be visualized in real-time in AR environment to demonstrate the mapping operation. The student can play with the physical piece to track the change through hand movement. For example, s/he can apply translation in any of the three directions of x, y and z through sliding the corresponding sliders; the transformed geometry will be augmented in AR environment; and the player can follow the translation action in the real environment by applying motion to the physical element (Figure1).

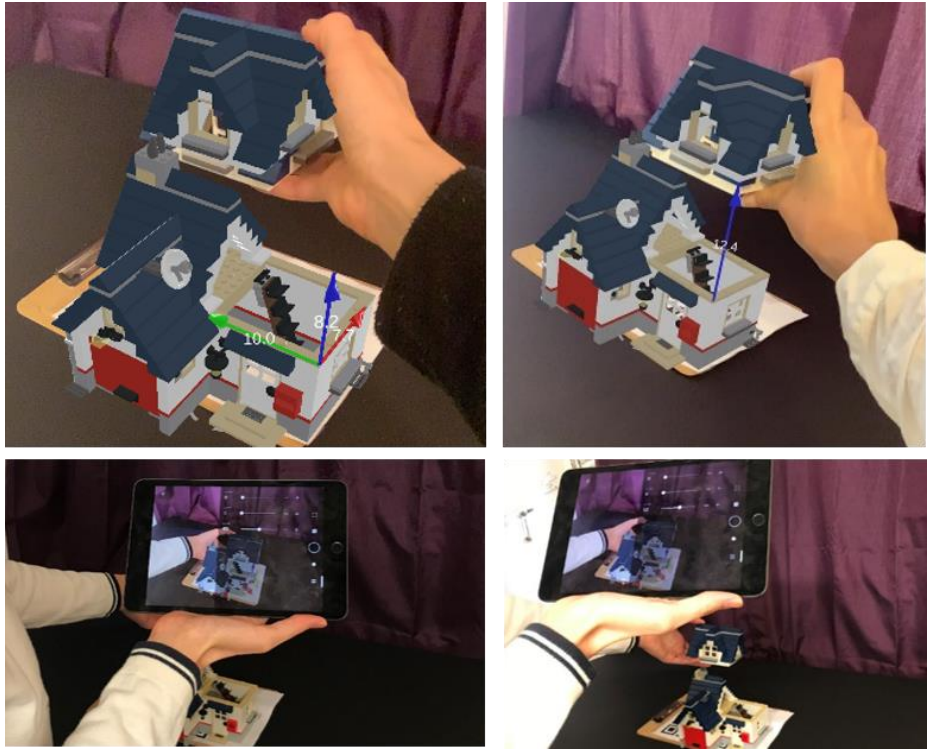

Figure 1: student plays with translation parameters in AR and tracks the transformation in physical environment

### 3.2 Prototype2

In this prototype a Python script is utilized to provide a customized rotation function using the RhinoScriptSyntax (RS) library. While the Grasshopper rotation component asks for two parameters (rotation angle and plane), the RS library needs three parameters i.e., rotation angle, axis and pivot. The function is customized by the authors to add the direction of the rotation as a separate parameter. The four parameters with their possible  parameter options/values are utilized as the input of the rotation function naming as: 1) rotation angle (-180º to 180º), 2) rotation axis (x, y, and z), 3) rotation direction (clockwise vs. counterclockwise), and 4) rotation base point (the local pivot point upon which rotation is applied). Figure 2 shows a snippet of the visual programming component and the corresponding parameters synched to Fologram AR.

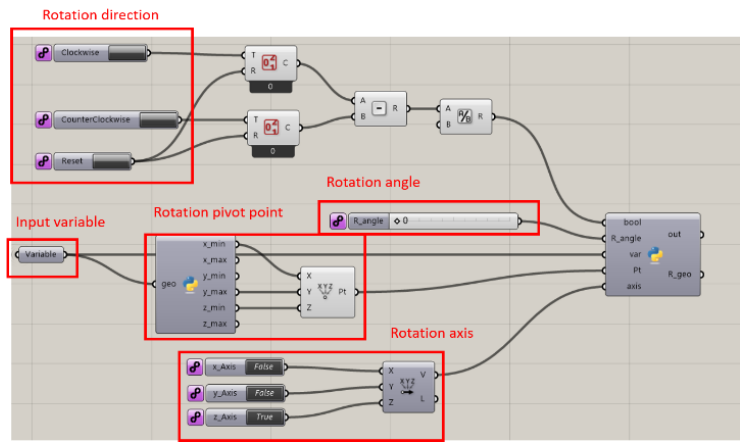

Figure 2: Grasshopper snippet for rotation function utilizing four input parameters and the geometry

In this prototype the variables are the rotatable openings (doors/windows) in the physical LEGO model. The player can choose which geometry s/he desires to apply rotation function by changing the slider that corresponds to the index of a selected geometry. Eight options exist based on the eight rotatable elements in the LEGO model. These elements could be rotated in the physical model with certain constraints, i.e. rotation axis, rotation direction and rotation pivot position. When the element is selected the position of the local pivot point upon which rotation will be applied is displayed as coordinate system (Figure 3).

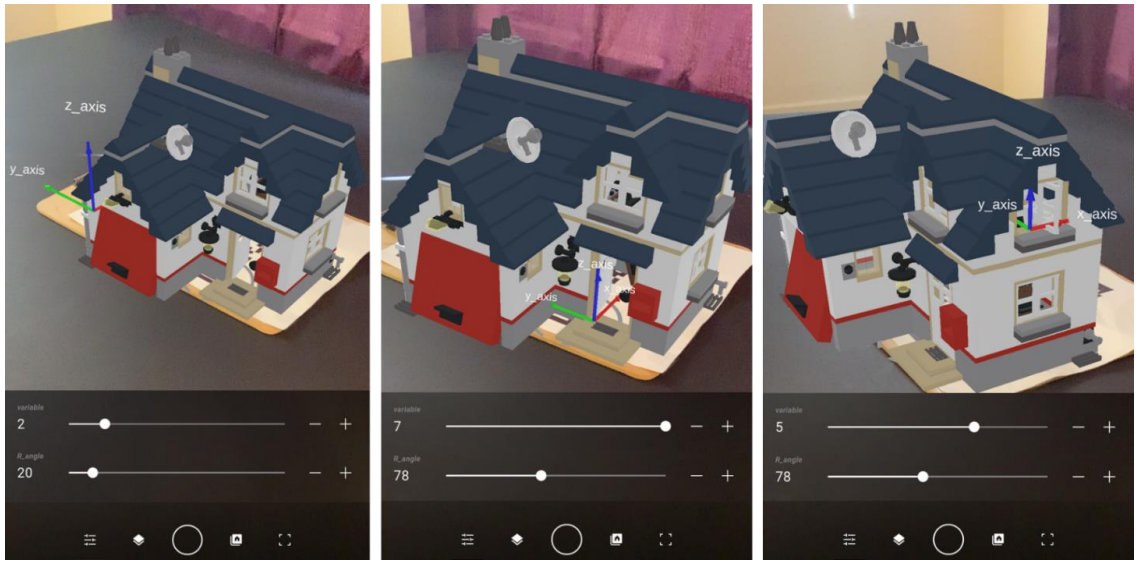

Figure 3. Player selects a modifiable opening (e.g., garage door, window, or entrance door) via slider representing variable indices to apply rotation using the angle slider.

The displayed coordinates may not be placed in a correct position of the selected geometry based on the applicable rotation of the physical element. Hence, students need to plug in the correct parameters to locate the pivot point on the correct position where rotation is applicable through the physical model. For this prototype the parameters to select the correct rotation point (six parameters shown in Figure 2) are accessible through Grasshopper, which is synched in real-time in the AR experiment. Later, students can play with other parameters such as rotation axis, rotation direction and angle displayed in AR through number sliders and/or toggle buttons (Figure 4).

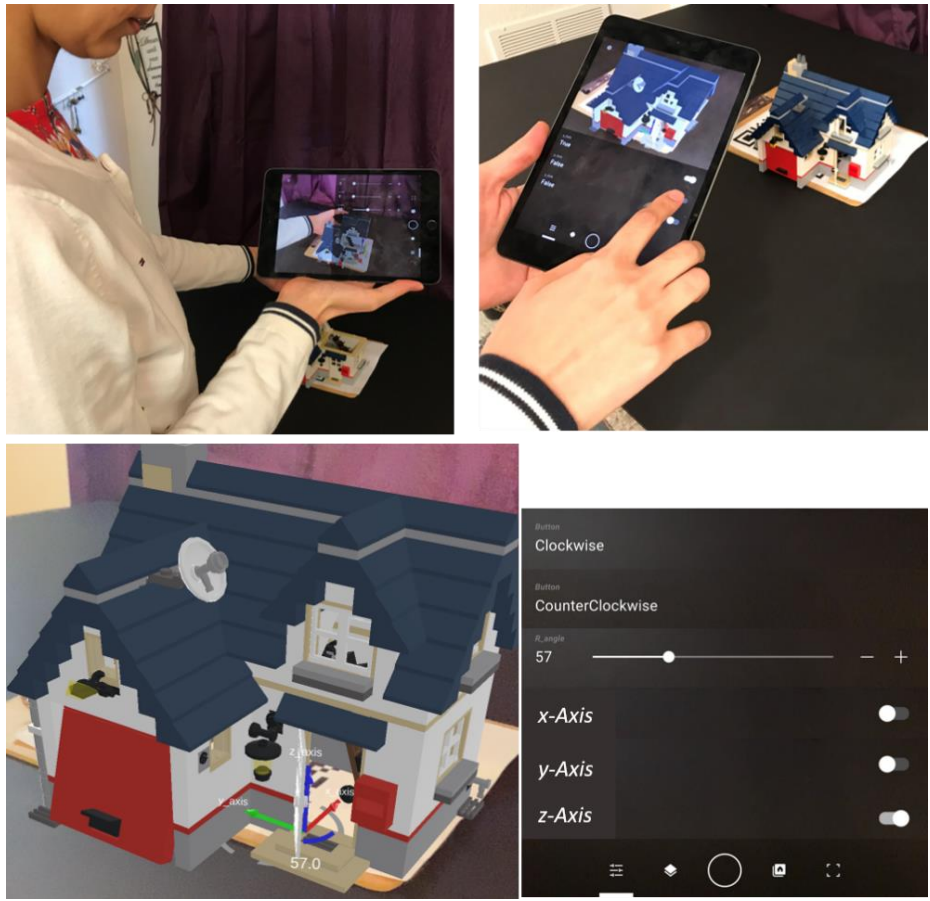

Figure 4. Student plays with different parameters of rotation function to match a transformation applicable in the corresponding physical piece. The bottom right image shows all potential parameters that player can modify through AR.

In this prototype the physical model is used not only to track the transformation in the physical world but also as a real time feedback to apply a transformation with specific parameters. Hence, although students can play with different parameters through various UIs and observe the augmented result, selecting a wrong parameter (e.g., wrong rotation axis, direction, or local pivot) results in a rotation which is not applicable in the physical model. For example, the entrance door can only rotate around z-axis in a clockwise direction when the pivot point is positioned at the left corner (Figure 5).

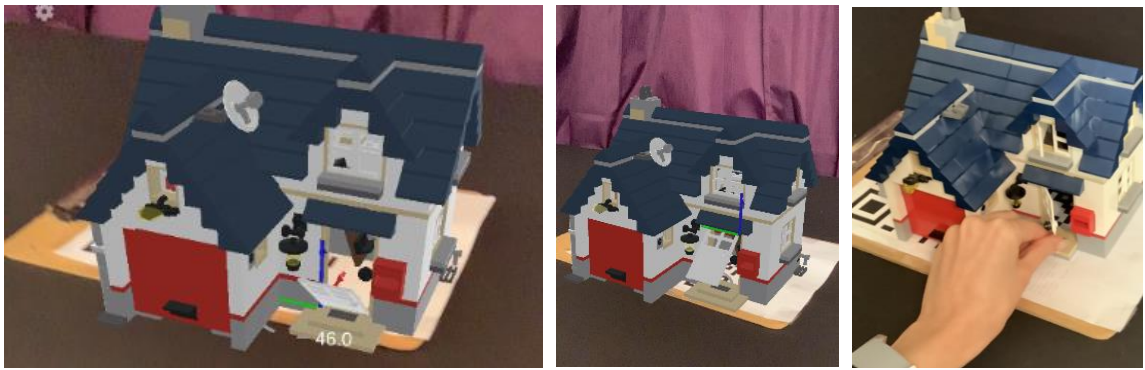

Figure 5. Student plays with different parameters to figure out the correct rotation parameters based on the physical model feedback. The left and middle images show rotations that are not feasible through the physical model. The right image shows the feasible rotation of the door in the physical LEGO model.

## 4  CONCLUSION AND FUTURE WORK

The study has presented an AR method for learning two common geometric transformations named translation and rotation as fundamental components of parametric modeling. In this study the geometric transformations are broken down into their corresponding components as input (variables and parameters) and operation functions to intuitively demonstrate geometric transformations as mappings and functions beyond motions in a "Learning Through Play" environment via a spatial experiment. The LEGO model is utilized as a physical tangible manipulative and a real-time feedback to improve spatial skill through hand movement; integrating physical and virtual object transformations with real-world experiment in an AR environment.

The application of this study could expand to other fields of design and engineering where creating a spatial experiment and contextualizing abstract information such as mathematical concepts related to geometry and transformations are significant in learning the subject. The current platform is limited to certain types of parameters in AR display (slider and button) and the communication is only one way, meaning that user needs to match the digital augmentation in the real world and the reverse automatic integration is not possible. Because the app is non-open source it has certain constraints for further customization. For example, it is limited in adding notations and graphical math representations.  In addition, hand and object occlusion is an issue in the current version of Fologram. Moreover, It is important to mention that working with Fologram in Grasshopper and conducting customizations may require knowledge and expertise in parametric modeling and using the required plugins. Considering the abovementioned constraints and limitations, the authors are developing a standalone educational app based on the assembly project – BRICKxAR [36] to address the challenges and issues exposed with Fologram AR in developing an educational platform for learning geometric transformations. In the future the authors intend to conduct user studies to evaluate the claim - AR enhancing learning of transformations - through experimental and control groups, using AR and non-AR learning methods, respectively. Multiple evaluation methods including spatial ability tests [37] , math tests, NASA_TLX [38], and motivation survey [33] will be used to assess the learning gain, cognitive load, motivation and engagement of students in learning the targeted subjects.

## ACKNOWLEDGEMENTS

The research is funded by the Texas A&M University's Presidential Transformational Teaching Grant (PTTG) and the Innovation [X] grant.